\documentclass[10pt]{article}

\begin{document}
\title{Excited Field of Particle in Quantum Theory}
\author{Gand Zhao \\ Department of Physics, Texas A\&M University, College Station, TX 77843}
\date{\today}
\maketitle

\begin{abstract}
A model about excited field of a particle is discussed.  We found
 this model will give wave-particle duality clearly and its
Lagrangian  is  consistent with Quantum Theory. A new
interpretation of quantum mechanics but not statistical
interpretation\cite{interpretation} is presented.
\end{abstract}

\section{Introduction}
Quantum Mechanics(QM) is one of the pillars of modern physics and
supported by a lot of precise experiment data. But, after one
century, the basic concept of QM, wave-particle duality is still
unclarified. The strangeness is, in Young's double slit
experiment, a photon has to go through both slits at the same time
to give interference pattern. To solve wave-particle duality, a
series of experiments were done, in which, the most important one
is the "Which way" experiment in an atom interferometer which was
proposed by Scully et al\cite{scully}, and successfully
realized by DNR\cite{dnr}.\\
In spite of trying to solve wave-particle duality experimentally,
we try to construct a more fundamental model which can give
wave-particle duality clearly and directly and be consistent with
Quantum Theory. Since to a particle there is wave-particle
duality, instead of made up by only field in Standard Quantum
Theory, we treat a particle as two parts, a point charge and its
excited field, where we assume all
 the interaction and energy-momentum of the
particle are taken by the excited field but not the point
charge(because of this, the point charge will disappear in
Lagrangian of particle and it is impossible for us to detect the
position of the point charge, we will show this in Sec.2). Before
we begin to construct the model, we first discuss this assumption
is
reasonable.\\
 In Classical Electromagnetic
Theory, when we calculate the force between two rest electrons A
and B, we first calculate the electric field excited by electron
A,

\begin{equation}
\overrightarrow{E}_A=\frac{e\overrightarrow{r}}{4\pi\varepsilon_0r^3}
\end{equation}
and the force between the two electrons is
\begin{equation}
|\overrightarrow{F}_{AB}|=Q_B|\overrightarrow{E}_{AB}|
\end{equation}
where $|\overrightarrow{E}_{AB}|$ is the amplitude of excited
electric field of electron A at the point of electron B. From the
the right hand of equation (2), we can see that its the excited
electric field of electron A which interacts with electron B. In
other word, if we treat the electron A as a system, including a
point charge and its excited electric
field, only the excited electric field interacts with electron B. \\
On the other hand, we know  electric field takes energy, and
electron B can also excite electric field, so we can calculate the
total energy of excited electric fields $\overrightarrow{E}_A$ and
$\overrightarrow{E}_B$
\begin{equation}
W=\frac{1}{2}\varepsilon_0\int{(\overrightarrow{E}_A+\overrightarrow{E}_B)^2dv} \\
=\frac{1}{4\pi\varepsilon_0}\frac{e^2}{R}+2C
\end{equation}
where R is distance between the two electrons, and C is constant
\begin{equation}
C=\frac{1}{2}\varepsilon_0\int{(\frac{e}{4\pi\varepsilon_0r^2})^2dv}
\end{equation}
We can calculate the gradient of energy W to get the force
\begin{equation}
\overrightarrow{F}=-\nabla{W}=\frac{e^2\overrightarrow{R}}{4\pi\varepsilon_0R^3}
\end{equation}
 It's consistent with Coulomb's law.  We can see that the
 interaction of two "rest" electrons could be totally expressed by
 the interaction of their excited electric fields

\begin{equation}
\overrightarrow{F}=-\frac{1}{2}\varepsilon_0\nabla{\int{d^3x2\overrightarrow{E}_A(x)\cdot\overrightarrow{E}_B(x)}}
\end{equation}

And from equation (3) ,(4) and (5) , the total energy(except the
mass which we will talk in Sec.2) of the two rest electrons are
also taken by their excited electric fields, but not the
point charge. \\
Here we have discussed that in Classical Electromagnetic Theory,
the interaction and energy of two rest electrons are taken by
their excited fields. Furthermore, we expect that, to a physical
particle (we separate it into two parts, a point charge and its
excited field), all interaction and energy-momentum are taken by
its excited field. Based on this argument, we can derive the
Lagrangian of excited field of a particle, and we will see it is
consistent with Quantum Theory. We can see, in this model, wave
particle duality is obvious, the point charge is particle like and
the excited field is wave like.
 \section{Model of Excited Field}
 In this section, we construct  model of excited field of a vector
particle, Scalar particle and Spinor particle, and we found its
Lagrangian is consistent with
Standard Quantum Theory. \\
Since a particle has wave-particle duality character, instead of
made up by only field, we treat a particle is made up by two
parts, a point charge and its excited field, and we assume all the
interaction and energy-momentum are taken by the excited field, as
argued in
previous section.\\
Recall that in Classical Electromagnetic Theory, a current of an
electron charge can excite electromagnetic potential
\begin{equation}
\partial_\nu{F^{\mu\nu}}=ej^\mu
\end{equation}
\begin{equation}
F_{\mu\nu}=\partial_\mu{A_\nu}-\partial_\nu{A_\mu}
\end{equation}
where $j_\mu=\rho_0U_\mu$, $\rho_0=\delta(\overrightarrow{x})$ is
rest electron charge density, $U_\mu$ is four velocity. \\
and the Lagrangian is
\begin{equation}
L=-\frac{1}{4}F^{\mu\nu}F_{\mu\nu}-ej^{\mu}A_\mu
\end{equation}
To a photon, in our model, is made up by two parts, a point charge
$e_B$ and its excited field $B_{\mu}$. Since photon is vector
particle, the excited field $B_{\mu}$ excited by $e_B$ should be
vector field, compared with (7), the equation of excited vector
field $B_{\mu}$  should be
\begin{equation}
\partial_\nu{F_B^{\mu\nu}}=e_Bj^\mu
\end{equation}
\begin{equation}
F_B^{\mu\nu}=\partial^\mu{B^\nu}-\partial^\nu{B^\mu}
\end{equation}
so, in our model, the Lagrangian of a free photon particle is
 \begin{eqnarray}
L=-\frac{1}{4}F_{B}^{\mu\nu}F_{B\mu\nu} -e_Bj^{\mu}B_{\mu}
\end{eqnarray}

where $j_{\mu}=\delta(\overrightarrow{x})U_{\mu}$. \\

We have to be very careful here. Different with Standard Quantum
Theory, in Lagrangian (12), there is no $-e_Bj^{\mu}B_{\mu}$ term,
and  to a free photon particle, the Lagrangian is simply
$L=-\frac{1}{4}F^{\mu\nu}F_{\mu\nu}$. While in our model, a
particle is made up by point charge $e_B$ and excited field
$B_\mu$,  the term $-e_Bj^{\mu}B_{\mu}$ is required to excite
vector field $B_\mu$. \\

We can write $B_{\mu}(x)$ as
 \begin{eqnarray}
B_{\mu}(x)=&&\int\frac{d^3\vec{p}}{(2\pi)^3}\int_{0}^{+\infty}dp_{0}e^{-i\vec{p}\cdot\vec{x}}(
B_{\mu}(p)e^{ip_0t}+B_{\mu}(p)e^{-ip_0t}) \nonumber \\
=&&B_{+\mu}(x)+B_{-\mu}(x)
\end{eqnarray}
where $B_{+\mu}(x)$ is positive frequency part and $B_{-\mu}(x)$
are negative frequency part.\\

In quantum field theory, causality requires  particle and
antiparticle have opposite quantum number\cite{peskin},  and only
the net excess of particle or antiparticle is
observable.\cite{s&ab}. While the positive frequency field is
corresponding to particle and negative frequency field is
corresponding to antiparticle, so the positive frequency field and
negative frequency field have opposite quantum number and only the
net excess of positive frequency field or negative frequency field
is observable, in other word, only the difference of positive
frequency and negative frequency field can appear in the
Lagrangian of interaction. We can write the Lagrangian of
interaction as

 \begin{eqnarray}
L=-f(B_{+\mu}-B_{-\mu},\partial{(B_{+\mu}-B_{-\mu})})V(x)
\end{eqnarray}

where $f(B_{+\mu}-B_{-\mu},\partial{(B_{+\mu}-B_{-\mu})})$ is
general function of $(B_{+\mu}-B_{-\mu})$ and its derivative,
$V(x)$ is other fields interacting with $B_{+\mu}$ and $B_{-\mu}$.
The
 exact form of
$f(B_{+\mu}-B_{-\mu},\partial{(B_{+\mu}-B_{-\mu})}) $ and $V(X)$
are given by gauge symmetry\cite{gauge}, here we don't talk about
this point.\\

The total Lagrangian becomes

 \begin{eqnarray}
L=-\frac{1}{4}F_{B}^{\mu\nu}F_{B\mu\nu} -e_Bj^{\mu}B_{\mu}
\nonumber \\
-f(B_{+\mu}-B_{-\mu},\partial{(B_{+\mu}-B_{-\mu})})V(x)
\end{eqnarray}

 Define $A_{\mu}\equiv{B_{+\mu}-B_{-\mu}}$,
then the Lagrangian becomes

\begin{eqnarray}
L=-f(A_{\mu},\partial{A_{\mu}})V(x)
-\frac{1}{4}F_B^{\mu\nu}F_{B\mu\nu} -e_Bj^{\mu}B_{\mu}
\end{eqnarray}
$A_\mu=B_{+\mu}-B_{-\mu}$ and $B_\mu=B_{+\mu}+B_{-\mu}$, they are
linear independent. But from (16), their is no energy and momentum
term of $A_{\mu}$. Since both $A_{\mu}$ and $B_{\mu}$ are vector
fields, they should have same form of energy and momentum , so the
Lagrangian is

\begin{eqnarray}
L=-\frac{1}{4}F_A^{\mu\nu}F_{A\mu\nu}-f(A_{\mu},\partial{A_{\mu}})V(x)\nonumber\\
-\frac{1}{4}F_B^{\mu\nu}F_{B\mu\nu} -e_Bj^{\mu}B_{\mu}
\end{eqnarray}
where
 \begin{eqnarray*}
F_{A\mu\nu}=\partial_{\nu}{A_{\mu}}-\partial_{\mu}{A_{\nu}}
\end{eqnarray*}
 We can see the first line of (17) is just the
Lagrangian of Maxwell field which interacts with other fields
$V(x)$. But now, $|A_{\mu(x)}|^2$ is not the amplitude of
possibility of a photon at point x any more. $A_{\mu}(x)$ is
$B_{+\mu}(x)-B_{-\mu}(x)$, where $B_{+\mu}(x)$ and $B_{-\mu}(x)$
are positive and negative frequency part of excited field of a
point charge. From the second line of (17), we can see the field
$A_{b\mu}(x)$ doesn't interact with any field( as for
$e_Bj^{\mu}$, the current of  point charge, there is no
interaction and energy-momentum, since all the interaction and
energy-momentum are taken by its excited fields $B_{+\mu}(x)$ and
$B_{-\mu}(x)$ , as assumed and argued in Sec.1). So if at initial
time the energy and momentum of $A_{b\mu}(x)$ is zero,
$P(E,\overrightarrow{P})=0$, at any later time, the
energy-momentum of $A_{b\mu}(x)$ is also zero,
$P^{'}(E,\overrightarrow{P})=0$. If at initial time, the energy
and momentum of $A_{b\mu}$ is not zero, at later time, if
$A_{\mu}$ interacts with other field, its momentum will change.
But  the momentum of $A_{b\mu}$ cannot change, $A_{\mu}$ and
$A_{b\mu}$ will be separated. We know $A_{\mu}(x)$ is
$B_{+\mu}(x)-B_{-\mu}(x)$ and $A_{b\mu}(x)$ is
$B_{+\mu}(x)+B_{-\mu}(x)$, where  $B_{+\mu}(x)$ and $B_{-\mu}(x)$
are two parts of excited field of a point charge $e_B$, $A_{\mu}$
and $A_{b\mu}$ have to be together, so the energy and momentum of
$A_{b\mu}$ must be zero. We can drop the second line of (17) into
background and what is left is

\begin{eqnarray}
L=-\frac{1}{4}F_{a}^{\mu\nu}F_{a\mu\nu}-f(A_{\mu},\partial{A_{\mu}})V(x)
\end{eqnarray}
this is consistent with Quantum Field Theory.(the exact form of
$f(A_{\mu},\partial{A_{\mu}})$ and $V(x)$ is given by
gauge symmetry\cite{gauge}) \\

Wave-particle duality appears at Lagrangian (17) clearly and
directly. The point charge $e_B$ is particle like and the
difference of positive and negative frequency of  its excited
field, $A_\mu$ is wave like. But since $e_Bj^{\mu}$, except of
exciting field $B_{\mu}$, doesn't interact with any field, it is
impossible for us to detect its position. In "Which way"
experiment\cite{scully}, according to our model, even we can tell
which slit the photon goes through, the interference pattern
is possibly not destroyed.  \\
In our model, both photon field $A_\mu(x)$ and electromagnetic
potential $A_\mu(x)$ in classical electromagnetic theory are
excited by a source. Electromagnetic potential is excited by
electron field, and photon field is excited by a point charge. So
$|A_{\mu}(x)|^2$ doesn't mean the possibility of a photon appears
at position x, as statistical interpretation.\cite{interpretation}
\\

 After constructing a model of a photon, we begin to construct
models of scalar particle and spinor particle.\\
 We know an electron charge density can excite scalar potential

 \begin{eqnarray}
\partial^{\mu}{\partial_{\mu}{\varphi}}=e\rho
\end{eqnarray}
 \begin{eqnarray*}
\rho=\delta(\overrightarrow{x})
\end{eqnarray*}

but obviously, $\varphi$ is not Lorentz invariant. To get an
equation of excited scalar field, we have to rewrite equation (17)
to be Lorentz invariant,

 \begin{equation}
\partial^{\mu}{\partial_{\mu}{\phi}}=\hat{e}\rho_0
\end{equation}
where $\phi$ is a scalar field, $\hat{e}$ is the charge to excite
$\phi$, and $\rho_0$ is Lorentz invariant charge "density", or
scalar charge density, we can write it as
 \begin{eqnarray*}
\rho_0=\frac{d\tau}{dt}\delta(\overrightarrow{x})
\end{eqnarray*}
To a point charge $e_{\phi}$ of a scalar particle and its excited
scalar field $\phi_B(x)$ where we separate $\phi_B(x)$ into
positive frequency part $\phi_{+}(x)$ and negative frequency part
$\phi_{-}(x)$, as same argument as Maxwell field, we can write the
Lagrangian as

 \begin{eqnarray}
L=\frac{1}{2}(\partial^{\mu}{\phi_B})^2
+e_{\phi}\rho_0\phi_B \nonumber \\
+f(\phi_{+}-\phi_{-},\partial{(\phi_{+}-\phi_{-})})V(x)
\end{eqnarray}

where $f(\phi_{+}-\phi_{-},\partial{(\phi_{+}-\phi_{-})})$
 is general function of $\phi_{+}-\phi_{-}$ and its
 derivative, V(x) is other fields which interacts with $\phi_{+}$ and
 $\phi_{-}$. \\
 Define $\phi\equiv{\phi_{+}-\phi_{-}}$ and put the
 energy-momentum term of $\phi$,
then the Lagrangian is

\begin{eqnarray}
L=\frac{1}{2}(\partial^{\mu}{\phi})^2+f(\phi,\partial{\phi})V(x)
\nonumber\\
 +\frac{1}{2}(\partial^{\mu}{\phi_B})^2
+e_{\phi}\rho_0\phi_B
\end{eqnarray}

where the first line of (22) is just the Lagrangian of scalar
field in QFT.

If $\phi$ interacts with Higgs field, then $\phi$ could be massive
scalar field, the  Lagrangian can be written as
\begin{eqnarray}
L=\frac{1}{2}((\partial^{\mu}{\phi})^2+m^2\phi^2)+f(\phi,\partial{\phi})
\nonumber\\
 +\frac{1}{2}(\partial^{\mu}{\phi_B})^2
+e_{\phi}\rho_0\phi_B
\end{eqnarray}

where m is the mass of $\phi$. \\
Since $\phi$ is massive but $\phi_b$ is massless($\phi_B$ cannot
interact with any field, including Higgs field), at the rest frame
of $\phi$, the energy and momentum of $\phi_B$ is zero, so at any
frame, the energy-momentum of $\phi_B$ is zero. Dropping the
second line of (23) into background( the energy and momentum of
$\rho_0$ is also zero, as argued in Sec.1), what is left is
\begin{eqnarray}
L=\frac{1}{2}((\partial^{\mu}{\phi})^2+m^2)+f(\phi,\partial{\phi})
\end{eqnarray}

which is just the Lagrangian of massive scalar field in Quantum
Field Theory. \\

Compared with equation (7) and (20), the equation of excited
spinor field should be like

 \begin{eqnarray}
\partial^{\mu}\partial_{\mu}{\psi}={\Lambda}{\hat{e}_{\psi}}\rho_0
\end{eqnarray}

where $\Lambda$ is a spinor number and $\bar{\Lambda}\Lambda=1$,
 $\rho_0=\frac{d\tau}{dt}\delta(\overrightarrow{x})$,
$\hat{e}_{\psi}$ is the charge to excite field $\psi$. \\
Define that
 \begin{eqnarray}
\partial_{\mu}{\psi}=\hat{g}_{\mu}(x)
\end{eqnarray}
 \begin{eqnarray}
\partial^{\mu}\hat{g}_{\mu}(x)={\Lambda}{\hat{e}_{\psi}}\rho_0
\end{eqnarray}

the Lagrangian is
 \begin{eqnarray}
L=-\bar{\psi}\gamma^{\mu}\frac{1}{i}\partial_{\mu}{\psi}+\bar{\psi}\gamma^{\mu}\frac{1}{i}\hat{g}_{\mu}+H.C
\end{eqnarray}

To a point charge $e_{\psi}$ of a spinor particle and its excited
spinor field $\psi_B$, where we separate $\psi_B(x)$ into two
parts, positive frequency $\psi_{+}(x)$ and negative frequency
$\psi_{-}(x)$, we can write the Lagrangian as
 \begin{eqnarray}
L=-\bar{\psi}_B\gamma^{\mu}\frac{1}{i}\partial_{\mu}{\psi_B}
+\bar\psi_B\gamma^{\mu}\frac{1}{i}g_{\mu}\nonumber\\
-f(\psi_{+}-\psi_{-},\bar\psi_{+}-\bar\psi_{-},\partial(\psi_{+}-\psi_{-}),
\partial(\bar\psi_{+}-\bar\psi_{-}))
V(x) \nonumber \\
+H.C
\end{eqnarray}
where
 \begin{eqnarray}
\partial^{\mu}g_{\mu}(x)={\Lambda}{e_{\psi}}\rho_0
\end{eqnarray}
and
$f(\psi_{+}-\psi_{-},\bar\psi_{+}-\bar\psi_{-},\partial(\psi_{+}-\psi_{-}),
\partial(\bar\psi_{+}-\bar\psi_{-}))$ is general function of $(\psi_{+}-\psi_{-})$,$(\bar\psi_{+}-\bar\psi_{-})$
 and their
derivatives. \\
 Here we don't write down the energy and momentum term of
$g_{\psi}(x)$ because it is not a real field, and  all  energy
momentum is taken by the excited field $\psi_{+}(x)$ and
$\psi_{-}(x)$ of point charge $e_{\psi}$.

Define $\psi=\psi_{+}-\psi_{-}$ and put the energy and momentum
term of $\psi$, then the Lagrangian becomes

\begin{eqnarray}
L=-\bar{\psi}\gamma^{\mu}\frac{1}{i}\partial_{\mu}{\psi}-f_1(\psi,\partial\psi)f_2(\bar\psi,\partial\bar{\psi})
V(x) \nonumber
\\-\bar{\psi}_b\gamma^{\mu}\frac{1}{i}\partial_{\mu}{\psi_b}
+\bar{\psi_b}\gamma^{\mu}\frac{1}{i}g_{\mu}\nonumber\\
+H.C
\end{eqnarray}

Dropping the second line of (31) and its H.C into background, as
argued as in (17) and (23), then we get

\begin{eqnarray}
L=-\bar{\psi}\gamma^{\mu}\frac{1}{i}\partial_{\mu}{\psi}-f(\psi,\bar\psi,\partial\psi,\partial\bar{\psi})
V(x)+H.C
\end{eqnarray}

 which is as same as the Lagrangian of spinor field in QFT. \\

If $\psi$ interacts with Higgs field, then $\psi$ could be massive
spinor field

\begin{eqnarray}
L=-\bar{\psi}\gamma^{\mu}\frac{1}{i}\partial_{\mu}{\psi}-m\bar\psi\psi-
f(\psi,\bar\psi,\partial\psi,\partial\bar{\psi})V(x)+H.C
\end{eqnarray}

So electron field $\psi$ can also be treated as the difference of
positive frequency and negative frequency part of excited field of
a point charge. We get such picture: a point charge $e_\psi$
 excites a field $\psi_B$, and we separate it into two parts:
positive frequency part $\psi_{+}$ and negative frequency part
$\psi_{-}$, but we found, all the interaction and energy-momentum
are taken by the difference of two parts of excited
field,$\psi=\psi_{+}-\psi_{-}$, or electron field, so we can only
"see" the electron field, as for the sum of the two parts of
excited field $\psi_B$ and the point charge $e_\psi$, they just
"disappeared". In other hand, electron field can excite electric
field. But the scale of excited electric field is much bigger than
the electron field, so in Classical Electromagnetic Theory, we can
treat the electron field as a "point charge".

\section{Discussion and Conclusion}
We have discussed a model of excited field of particle. In this
model, to a particle, there is two parts, a point charge and its
excited field. And in this this model, wave particle duality is
very clearly, point charge is particle like and excited field is
wave like. But, as seen in Lagrangian (17),(23) and (31), the
Lagrangian of  point charge can be dropped into background, and it
is impossible to detect its position. After dropping background ,
we get Lagrangian (18),(24) and (32), and they are consistent with
QFT.
 But now, to a field
$\chi(x)$, $|\chi(x)|^2$ doesn't mean the amplitude of possibility
of a particle appearing at point x. $\chi(x)$ is just present the
difference of positive and negative frequency part of excited
field by a point charge (or current). So we get same Lagrangian(if
we drop the background) as QFT, but the interpretation are totally
different. If this model is correct, then we have to reconsider
some parts of
quantum physics, especially  string theory . \\
The main difference of this model and standard Quantum Theory in
Lagrangian is that there is unusual background, as shown in (17),
(23) and (32). With more study about this background, possibly we
can testify
this model in experiment.\\
We have to mention that based on the argument that both
interaction and energy-momentum of a particle are taken by its
excited fields, the model we've discussed is not unique to satisfy
Quantum Theory.

\end{document}